# Separation of Body and Background in Radiological Images. A Practical Python Code

**Seyedeh Fahimeh Hosseini [1], Faezeh Shalbafzadeh[2,3], Behzad Amanpour-Gharaei [3,4]***

[1] Medical Physics Department, School of Medicine, Iran University of Medical Sciences, Tehran, Iran.
[2] Department of Medical Physics and Biomedical Engineering, School of Medicine, Tehran University of Medical Sciences, Tehran, Iran.
[3] Research Center for Molecular and Cellular Imaging, Advanced Medical Technologies and Equipment Institute, Tehran University of Medical Sciences, Tehran, Iran.
[4] Cancer Biology Research Center, Cancer Institute, Tehran University of Medical Sciences, Tehran, Iran.

*Corresponding Email: behzad.amanpour@gmail.com

## Abstract

Radiological images, such as magnetic resonance imaging (MRI) and computed tomography (CT) images, typically consist of a body part and a dark background. For many analyses, it is necessary to separate the body part from the background. In this article, we present a Python code designed to separate body and background regions in 2D and 3D radiological images. We tested the algorithm on various MRI and CT images of different body parts, including the brain, neck, and abdominal regions. Additionally, we introduced a method for intensity normalization and outlier restriction, adjusted for data conversion into 8-bit unsigned integer (UINT8) format, and examined its effects on body-background separation. Our Python code is available for use with proper citation.

## 1. Introduction

Radiological images, including magnetic resonance imaging (MRI) and computed tomography (CT) images, usually have a dark background that shows the area outside of patients' bodies. For many analyses, it is necessary to separate the image background from the body parts [1-4]. These analyses range from simple tasks, such as calculating the average intensity of an image, to more advanced ones, like skull stripping of brain images [3, 5] and segmentation of body parts [2]. By separating body parts from the background, we can achieve several advantages, such as improving data compression, image readability, diagnostic performance, and defining cost functions.

Apart from deep learning algorithms, there are several methods for separating body parts from the background and performing background removal, including thresholding, edge detection, contouring, and morphological operations [1-5]. Various programming languages can be used to implement these algorithms, such as MATLAB [6] and Python [7]. We utilized Python programming, which offers advantages over other languages in terms of availability, development, user communities, and application in many industries. The primary library used in our program is the Open Source Computer Vision Library (OpenCV) for Python, which is open-source and includes numerous computer vision algorithms [7].

In this article, we introduce an unsupervised Python algorithm for separating the body part from the background of radiological images. We tested the algorithm on a variety of MRI and CT images from different parts of the body, including the brain, neck, and abdominal regions. Its code is available for readers to test and utilize with proper citation. We designed our algorithm as a Python function that accepts multiple inputs and returns a binary mask that segments the body part. Each input to the function alters the output mask, providing flexibility to handle different conditions in images. Additionally, we introduce a normalization function that is modified to be used before converting the data type of an image to 8-bit unsigned integer (UINT8, range: 0 to 255 decimal), as the OpenCV functions of our algorithm operate only on the UINT8 data format. The remainder of the paper presents different parts of our algorithm in a practical manner, making it accessible for use, and we have plotted the outputs in several figures. The contributions of this paper can be outlined as follows:

1. Introducing an easy-to-handle Python program for separating human body from the background of radiological images.
2. Introducing an image normalization algorithm that both normalizes and restricts outlier values of an image for converting its data type to UINT8.
3. Making the Python code available for use with proper citation (the tested images may also be shared upon reasonable request).

## 2. Material and Methods

### 2.1. Data

The 2D images used in this study are a set of fully anonymized MRI and CT images acquired at different imaging centers in Tehran, Iran, and have been utilized in our previous studies. The study was conducted in accordance with the Ethics Committee of Tehran University of Medical Sciences, and the requirement for informed consent was waived since the images are single slices and anonymized. We also tested our algorithm on an anonymized 3D MRI image. However, we cannot share this 3D image because reconstructing the patient's face might be possible.

### 2.2. Image Normalization for UINT8

Intensity normalization is an important image pre-processing step that reduces inhomogeneities within an image and between images from different sources. This process has applications in many medical image analyses, including classification, segmentation, imputation, and registration [8].

There are several image intensity normalization algorithms [8]. We named our normalization function "*NormalizeForUINT8_OutlierRemove*". First, it calculates the z-score of its input image, commonly referred to as "standardization," because it does not restrict the intensity values of the image to a certain range [9]. To calculate the z-score of an image, we subtract the mean pixel intensity of the image from each pixel intensity value, and then divide the result by the standard deviation of the pixel intensities. This results in an image with a mean value of zero and a standard deviation of one. The pixel intensity values can range from minus infinity (-∞) to plus infinity (+∞), but typically around 99% of values lie between -3 and 3 [9]. Therefore, by default, we considered pixel values outside this range as outliers. We provided an input for the normalization function called "*limit*" that could be adjusted by the user. Pixel values outside the range [-limit, +limit] are considered outliers and are restricted to the upper and lower limits. Finally, the range [-limit, +limit] is mapped to [0, 255], which is UINT8, and is required for the main function of the algorithm.

### 2.3. Background Body Separation

The primary function of our algorithm utilizes a combination of Python libraries, including OpenCV (version 4.10.0), SciPy (version 1.13.1), NumPy (version 1.26.4), and Matplotlib (version 3.7.1). Below is a summary of the inputs and output, which are explained in detail afterward.

Inputs (the gray ones are the function hyperparameters):
- image: The input image (2D array).
- *normalization* (default = 'OFF'): Indicates whether to use our normalization function.
- *limit* (default = 3): The outlier limit for the normalization function.

- *contour_number* (optional): Specifies a particular contour in the image for drawing the mask.
- *thickness* (optional): The thickness of the contours to be drawn.
- *plot* (default = 'ON'): Indicates whether to show some plots.
- *vmin* (optional): The minimum value of the colormap range.
- *vmax* (optional): The maximum value of the colormap range.

Output:
- mask: The binary mask that separates the patient's body from the image background.

The first step of our main function is converting the image data type into UINT8 because some OpenCV functions only work with this data type. If the user sets *normalization = 'ON'*, the main function utilizes the normalization function we described in the previous section, which also converts the image data type into UINT8. Otherwise, the image data type is converted into UINT8 using "numpy.uint8" from the NumPy library.

The second step is image thresholding using the OpenCV thresholding function [10]. Thresholding is a binary segmentation that compares pixel values with a threshold value, setting values smaller than the threshold to 0 and the rest to a maximum value. We used Otsu's method, provided by OpenCV, which determines an optimal threshold value from the image histogram [10].

The next step is finding all possible contours in the image and filling them to generate an initial mask. A contour is a curve joining all the continuous points along the boundary of an object in an image that have the same color or intensity. We used the "findContours" function from OpenCV to find as many contours as possible in the image. Afterward, we used another OpenCV function called "drawContours" to draw the contours and fill inside them to produce the initial mask. The user can use another input of the main function called *contour_number* to shape the mask based on a specific contour instead of all possible contours. *contour_number = 1* means the final mask will be based on the largest contour in the image, *contour_number = 2* means the second largest contour, etc. The final step is filling the remaining holes in the initial mask to generate the final mask. For this, we used the "ndimage.binary_fill_holes" function from the SciPy library of Python.

After generating the final mask, the main function, by default, displays three plots: the input image, the UINT8 version of the input image, and the final mask. The "matplotlib.pyplot.imshow" and "matplotlib.pyplot.subplots" functions are used to show the plots. If the user does not want to see the plots, they can set *plot = 'OFF'*. Matplotlib automatically assigns a colormap range to the pixel values of the input image. For grayscale images, this range is from black to white. The user can set a desired intensity range using *vmin* and *vmax*. Pixels with values <= *vmin* will be completely black, and those with values >= *vmax* will be completely white.

# 3. Results

## 3.1. Effect of Normalization

Fig. 1 shows the effect of using our function, *NormalizeForUINT8_OutlierRemove*, on the appearance of the input image after converting it to the UINT8 data format. Using this function does not necessarily improve the output mask.

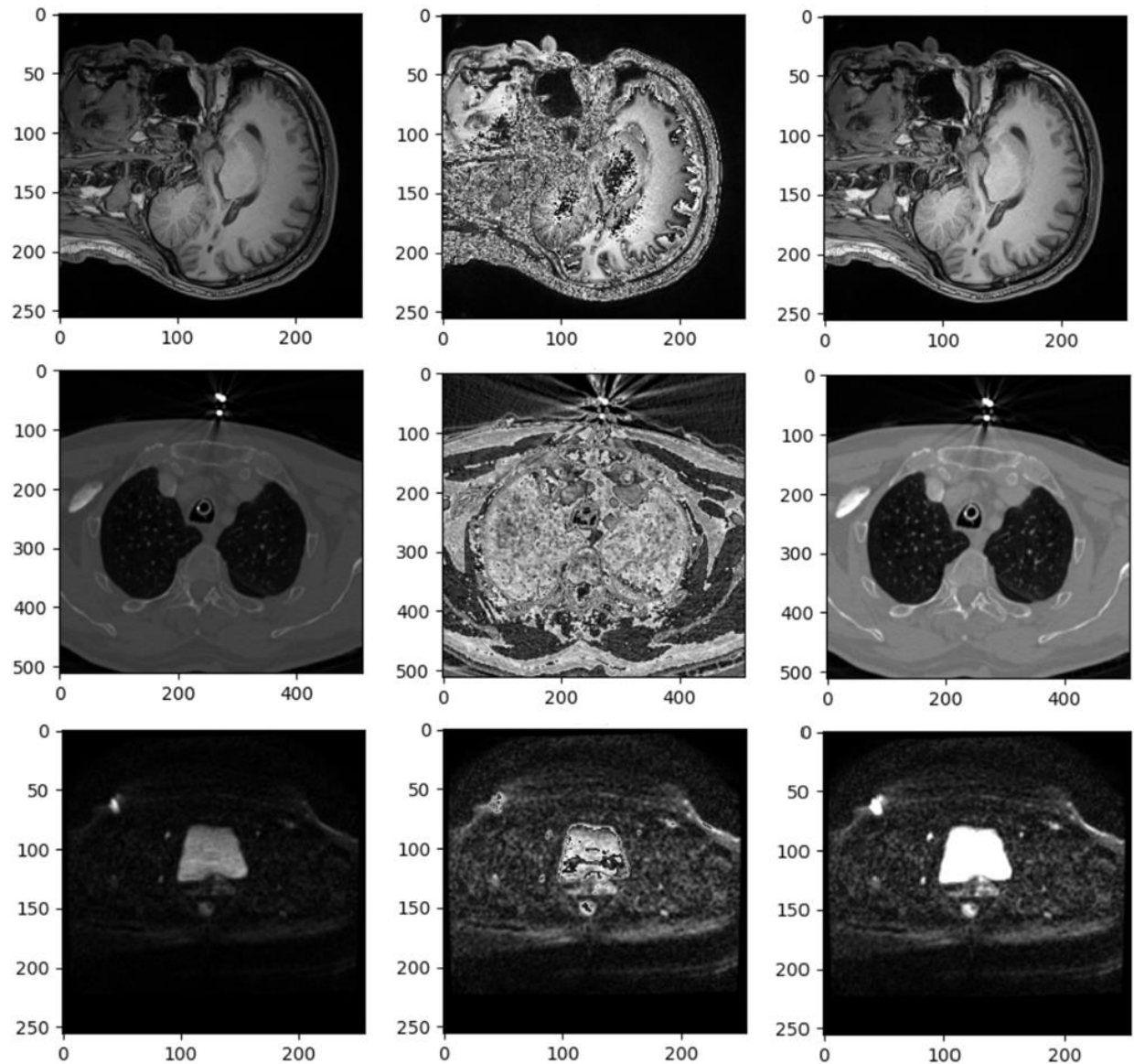

**Figure 1.** Effect of using the function *NormalizeForUINT8_OutlierRemove* on the appearance of the input image after converting to the UINT8 data format. The left column shows the input images. The middle column shows the input images after converting to the UINT8 data format without normalization (default choice). The right column shows the normalized input images after converting to the UINT8 data format.

## 3.2. Background Body Separation

Fig. 2 shows the final generated masks for most of images tested in this study (the rest are in Fig. 7). The shape of the masks can be changed by adjusting the main function input values.

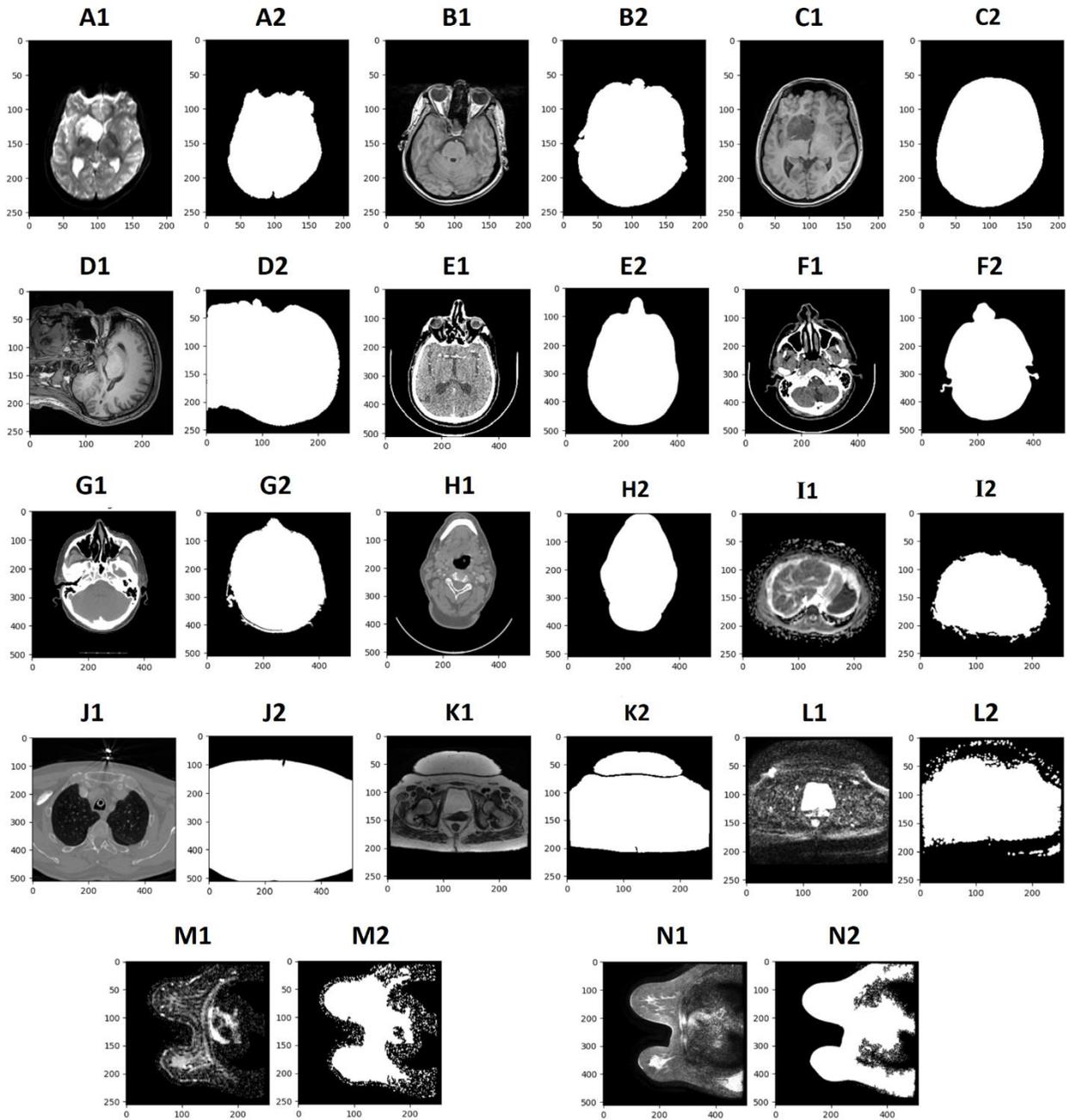

**Figure 2.** Set of input images (A1-N1) and their corresponding masks (A2-N2). A) Diffusion-weighted imaging (DWI), axial brain slice. B & C) Brain T1 MRI, axial slices. D) Brain T1 MRI, sagittal slice. E) Brain computed tomography (CT) image, axial slice. F & G) Skull base CT, axial slice. H) Neck CT, axial slice. I) Liver apparent diffusion coefficient (ADC) map, axial slice. J) Chest CT, axial slice. K) Pelvis T2 MRI, axial slice. L) Pelvis DWI B0 image, axial slice. M) Breast ADC map, axial slice. N) Breast T1 MRI, axial slice.

Table 1 demonstrates the characteristics of all images tested in this study (Fig. 2 and 7) and the input values (hyperparameters) adjusted to achieve the best possible mask that separates each image.

**Table 1:** The characteristics of the input images and their corresponding input values (hyperparameter values) for the main function.
"-" indicates the default value for the corresponding input.

| Appearance in Article | Image Type | Data format | Min-Max values | *normalization* | *limit* | *thickness* | *contour number* |
|---|---|---|---|---|---|---|---|
| Fig. 2. A | DWI, Brain | uint8 | 0-254 | - | - | - | - |
| Fig. 2. B | T1 MRI, Brain | uint8 | 0-254 | - | - | 3 | - |
| Fig. 2. C | T1 MRI, Brain | uint8 | 0-254 | - | - | - | - |
| Fig. 2. D | T1 MRI, Brain | float64 | 0.0-936.0 | - | - | 3 | - |
| Fig. 2. E | CT, Brain | uint16 | 0-2784 | ON | - | - | 1 |
| Fig. 2. F | CT, Skull_base | uint16 | 0-2923 | ON | - | - | 1 |
| Fig. 2. G | CT, Skull_base | int16 | -2000-3172 | - | - | - | 1 |
| Fig. 2. H,  Fig. 7. D | CT, Neck | uint16 | 0-2766 | ON | - | - | 1 |
| Fig. 2. I,  Fig. 7. B | ADC map, Liver | uint8 | 0-255 | - | - | - | 1 |
| Fig. 2. J,  Fig. 7. C | CT, Chest | uint16 | 0-4095 | ON | - | - | 1 |
| Fig. 2. K | T2 MRI, Pelvis | int16 | 0-1668 | ON | - | - | - |
| Fig. 2. L,  Fig. 7. A | DWI_B0, Pelvis | int16 | 0-983 | ON | 0.5 | 2 | - |
| Fig. 2. M | ADC map, Breast | uint8 | 0-253 | ON | 1 | - | - |
| Fig. 2. N | T1 MRI, Breast | uint8 | 0-255 | ON | 1 | - | - |
| Fig. 7. E | CT, Neck-Teeth | int16 | -2000-4095 | - | - | - | 1 |
| Fig. 7. F | CT, Brain | int16 | -2000-2844 | - | - | - | 1 |

### 3.3. Effect of Contour Thickness
Fig. 3 shows the effect of varying *thickness* values on the generated mask for some images. Higher thickness values present both advantages and disadvantages, which are explained in the discussion section.

### 3.4. Specific Contours
While the primary functionality of the code is to separate the body from the background, specific contours can be achieved by adjusting the main function input values (hyperparameters), particularly the input called *contour_number*. Fig. 4 provides examples of such specific contours.

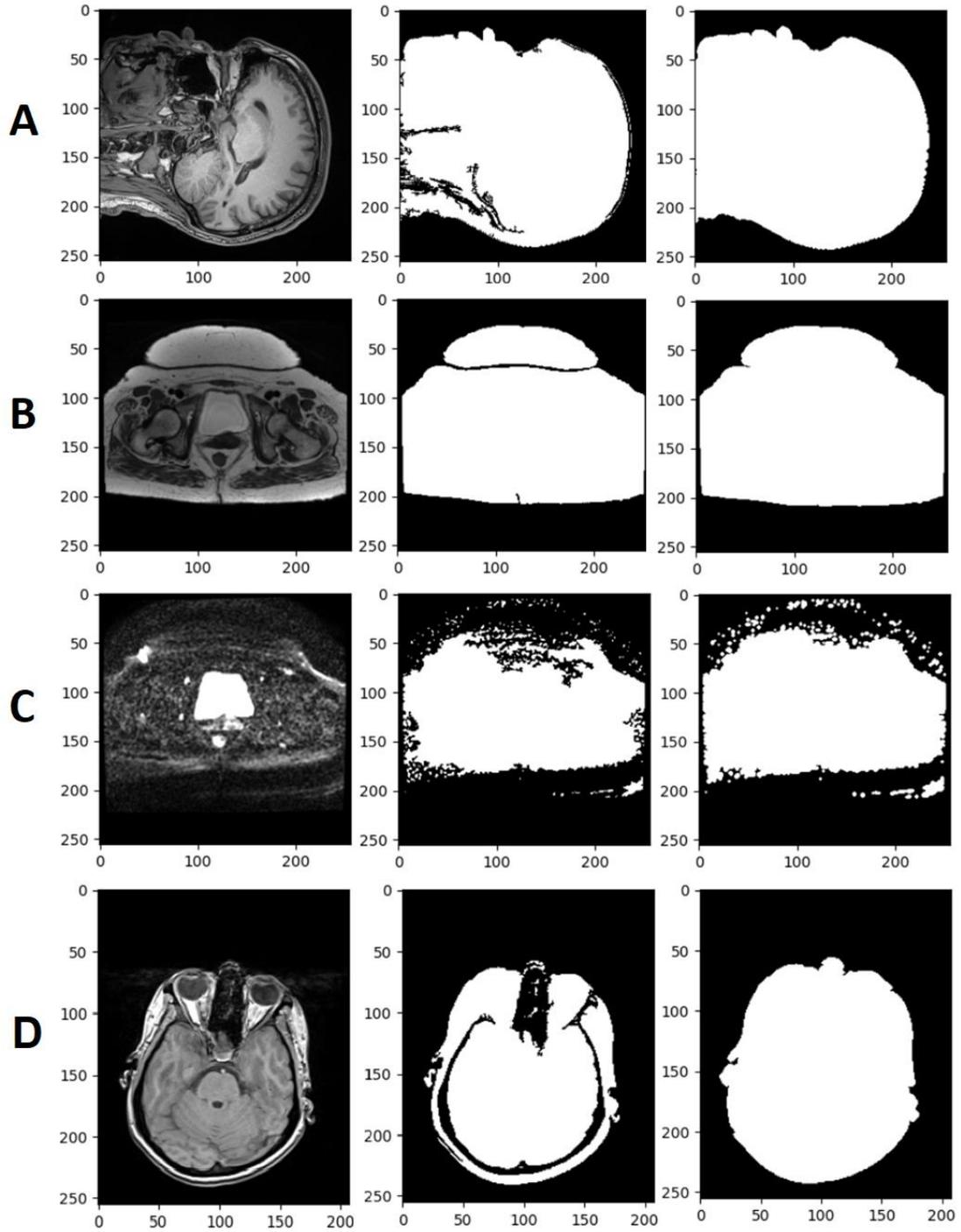

**Figure 3.** Effect of thickness values on the mask. The left column displays the input images. The middle column displays the generated masks for the images using the default value (*thickness = -1*). The right column displays the generated masks with different *thickness* values: A & B) *thickness = 2*, C & D) *thickness = 3*.

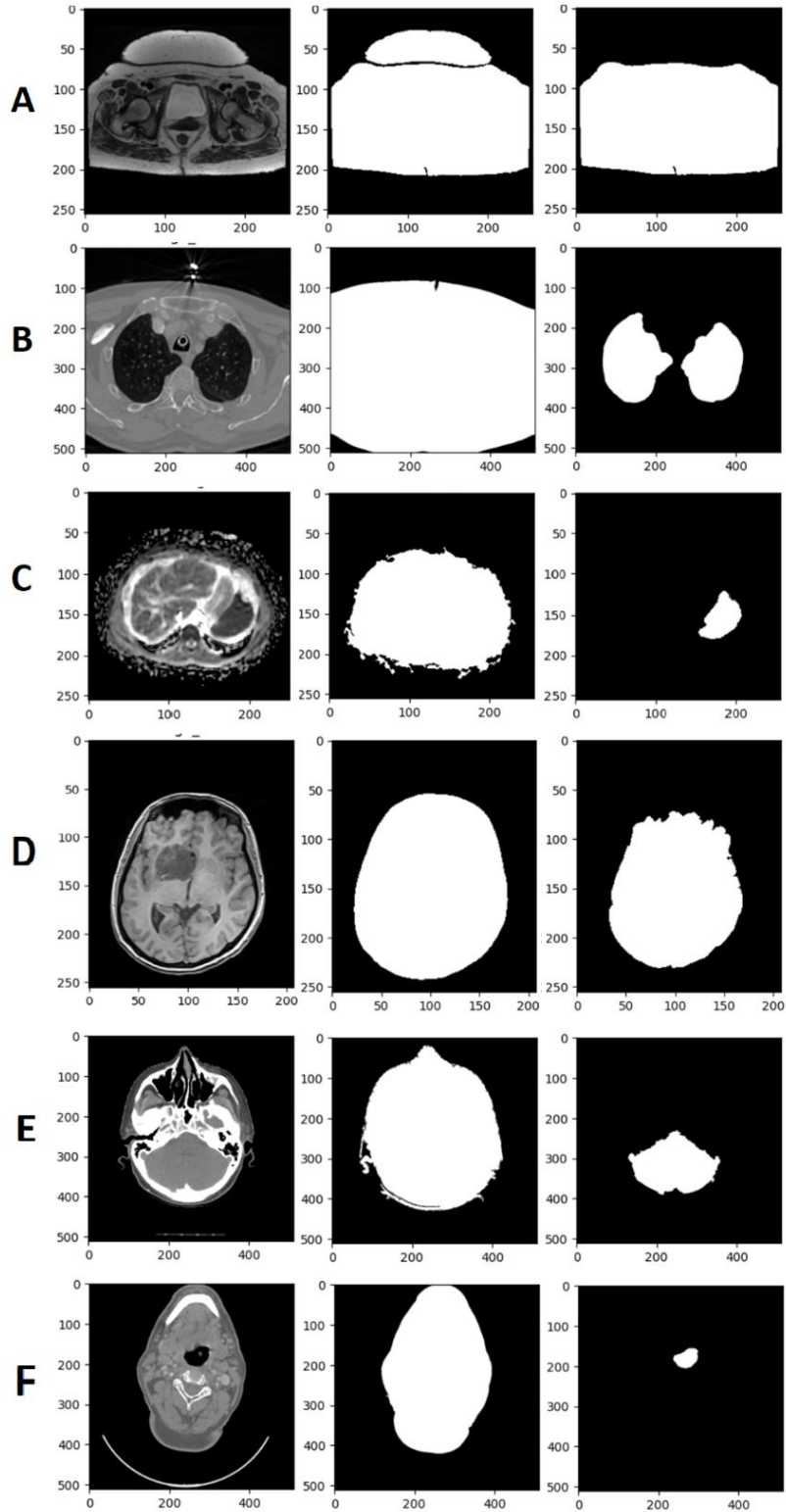

**Figure. 4.** Specific contours generated by the main function. The left column shows the input images. The middle column shows the mask that separates the body from the background. The right column demonstrates how the mask can be altered by selecting: A) *normalization='ON'*, *contour_number=1*, B) *normalization='ON'*, *contour_number=2* and *contour_number=3*. Each of the contours is achieved separately, and their masks should be summed up.
C) *contour_number=2*, D) *contour_number=3*, E) *contour_number=2*, F) *normalization='ON'*, *contour_number=2*.

## 3.5. Effect of Outlier Limit

Fig. 5 illustrates the impact of various *limit* values on the final mask produced by our function. Lower *limit* values result in a denser mask.

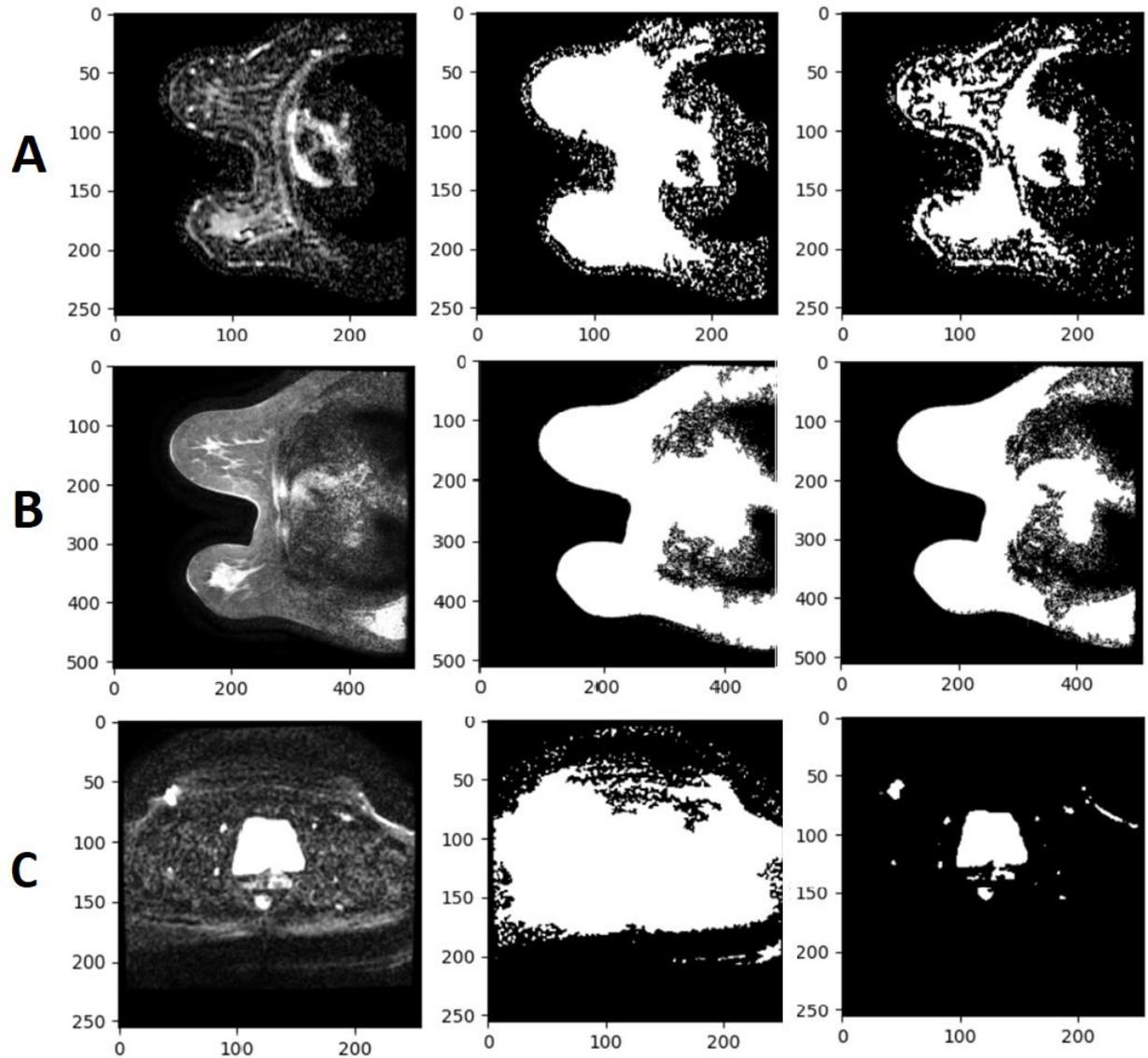

**Figure 5.** Effect of Outlier Limit in Normalization. The left column shows three input images. The middle column shows the final mask after normalization with A & B) *limit = 1* and C) *limit = 0.5*. The right column shows the final mask after normalization with the default value of *limit = 3*.

## 3.6. Effect of *vmin* and *vmax* on Input Image Appearance

Fig. 6 demonstrates how the values of *vmin* and *vmax* could alter the appearance of the input image. These values do not affect the final mask.

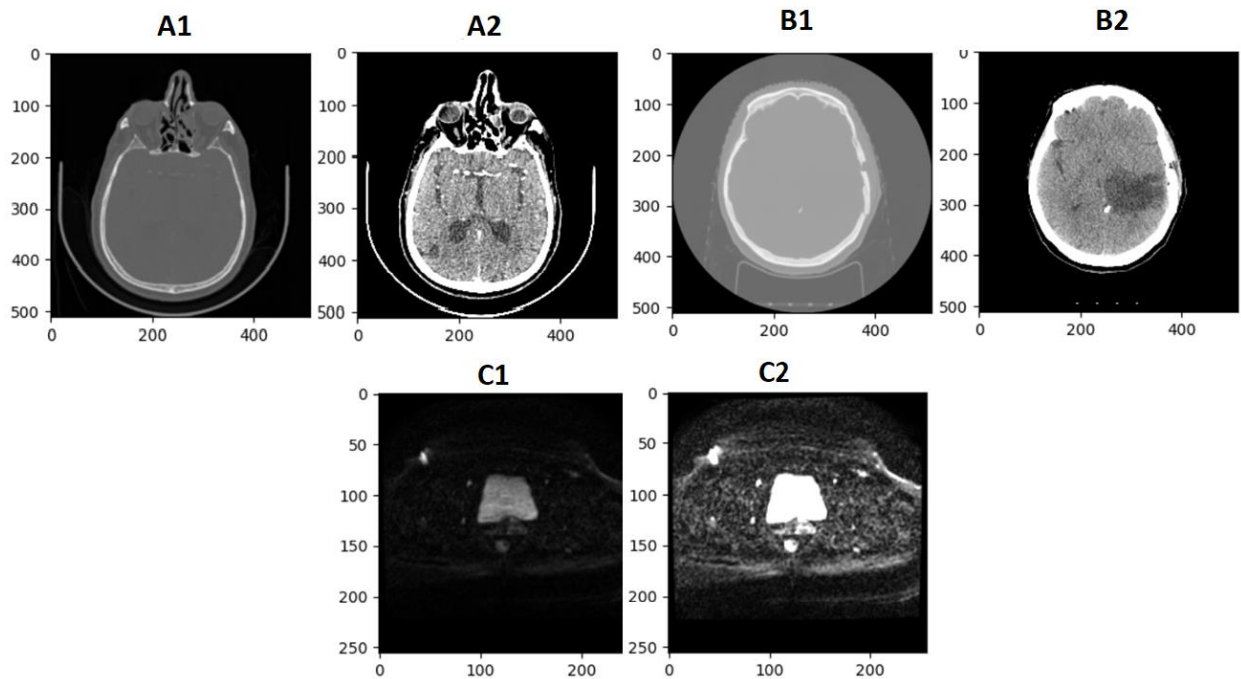

**Figure 6.** Effect of *vmin* and *vmax* on Input Image Appearance. The first row shows the default appearance of two brain CT images (A1 & B1) using the Matplotlib library, and their adjusted appearances with *vmin = 1000* and *vmax = 1100* (A2 & B2). The second row shows the default appearance of a Pelvis DWI image (C1), and with *vmin = 0* and *vmax = 200* (C2).

## 3.7. Image with Artifacts

Fig. 7 illustrates the effects of background artifacts on the output masks for two MRI and four CT images. These artifacts have influenced the shape of some masks, which is further discussed in the discussion section.

## 3.8. 3D Mask

The input image for the main function should be a 2D image. We created a 3D mask for a 3D image by applying the function to its 2D slices individually. Fig. 8 shows some slices of the 3D image, their corresponding masks, and the final 3D mask.

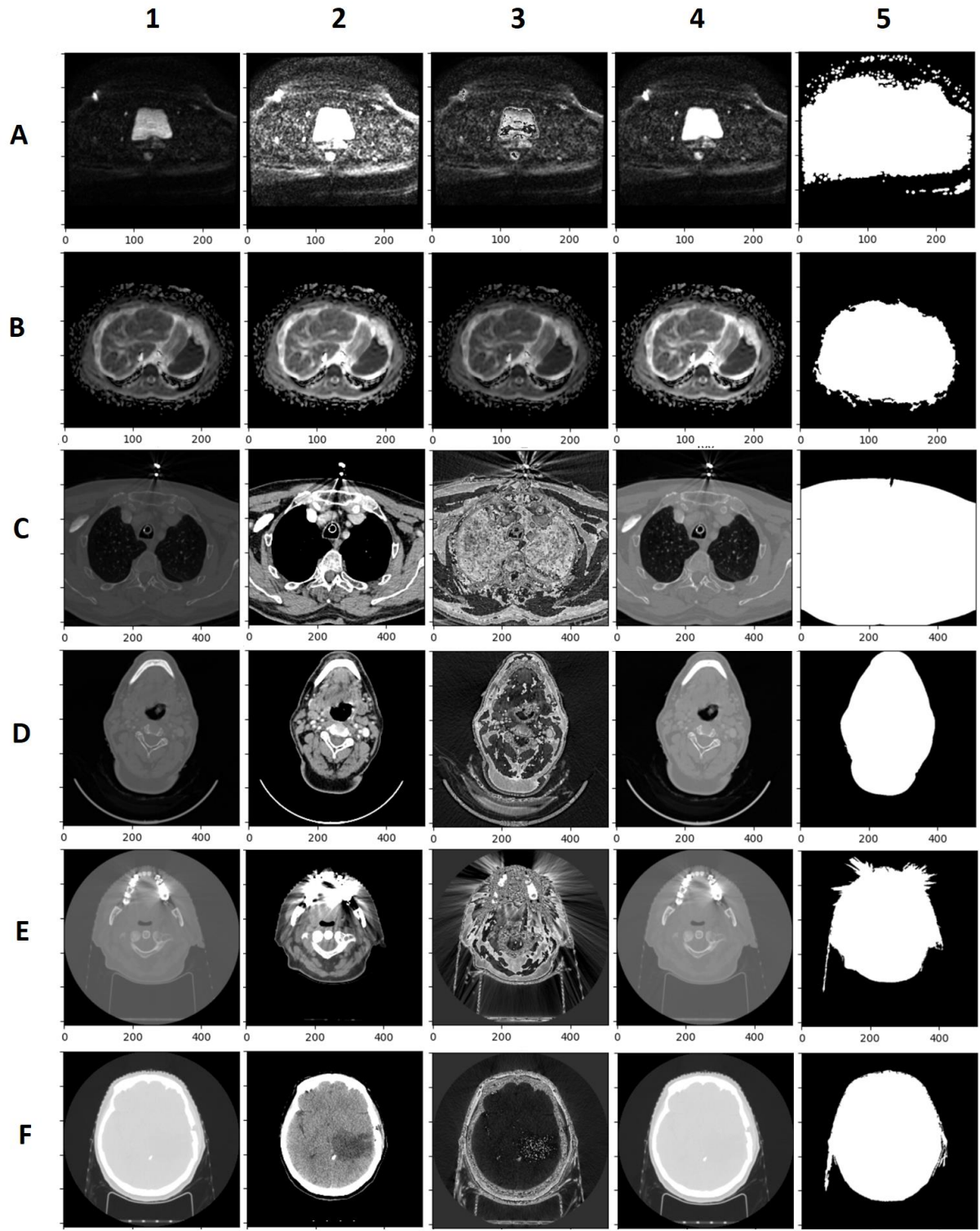

**Figure 7.** Effect of Background Artifacts on the Generated Masks.
The columns show transformations on input images, including: 1) Input image, 2) Input image with adjusted brightness and contrast using *vmin* and *vmax*, 3) Input image converted to UINT8 data format without normalization, 4) Input image converted to UINT8 data format after normalization, 5) Final mask.
The rows display images of different body parts with background artifacts, including: A) Pelvis DWI_B0, B) Liver ADC map, C) Chest CT, D) Neck CT, E) Neck-Teeth CT, F) Brain CT.

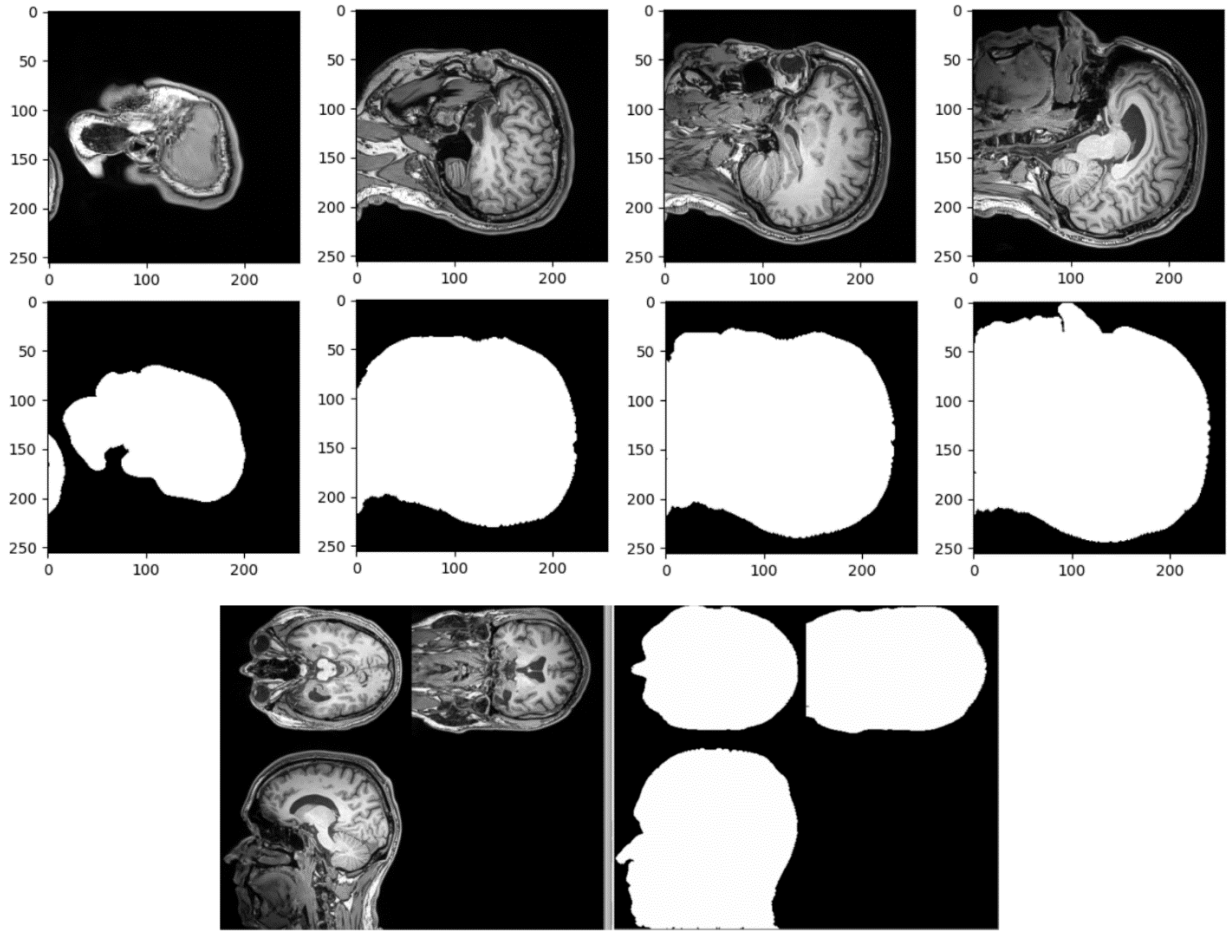

**Figure 8.** Making a 3D Body Mask. The first two rows show sagittal slices of a 3D brain MRI image and their corresponding body (head) masks. By joining these masks, we formed a 3D mask, which is shown in the bottom row.

## 4. Discussion

### 4.1. Normalization for UINT8 and Outlier Limit

As shown in Fig. 1, the normalization function enhances the appearance of the image after converting it to the UINT8 data format. However, according to the column *normalization* in Table 1, it improved the final mask for only 50% of the images we tested. Therefore, it is not mandatory for separating body and background regions in radiological images. Nevertheless, we need to test a larger set of images for a more conclusive result.

An advantage of using the function *NormalizeForUINT8_OutlierRemove* is the ability to control outlier restriction through the hyperparameter *limit*. The lower the limit, the more pixel values of the image are considered outliers and are restricted, resulting in a denser mask. As shown in Fig. 5, this could be beneficial for DWI images and breast images.

### 4.2. Contour Thickness

While our algorithm attempts to fill the holes within its generated mask, there may still be holes and empty areas inside the mask, as shown in Fig. 3. In this case, higher values for the parameter

*thickness* could help fill these empty areas. However, a larger thickness will result in a larger mask than the body, occupying the background pixels around the body borders (e.g., larger head, larger nose, etc.).

### 4.3. Background Artifacts

Fig. 7 shows some possible background artifacts in DWI (MRI) and CT images. In DWI and ADC map images (Fig. 7, first two rows), where each pixel intensity correlates with the diffusion of water molecules at that point, the body border is not clear and has discontinuous bright and dark points. This causes distortion in the mask's outer border.

In CT images, background artifacts are common. Fig. 7.C shows a jewelry artifact in front of the patient's chest that is removed in the output mask. Fig. 7.D and 7.F show artifacts due to physical objects around the patients (table artifacts), which are removed in Figure 7D but not completely in Figure 7F.

In Fig. 7.E, there are two types of CT artifacts. One is a dental beam streak artifact due to teeth with metal fillings. The other is a table artifact. The best final mask we achieved by adjusting the hyperparameters still shows parts of the artifacts. This is a limitation of our work that we plan to improve.

## Ethics Statement

All data were provided anonymously, and no information about the patients' diseases or histories was shared.

## Data Availability

- The Python code is available for use with proper citation at:

https://github.com/Behzad-Amanpour/medical_image_pre-processing/tree/main/Body_Background_Separation

- All 2D images can be accessed from the corresponding author upon reasonable request. The 3D image cannot be shared because the reconstruction of the patient's face might be possible.